\documentclass[12pt]{article}

\usepackage{amsmath}
\usepackage{amssymb}
\usepackage{verbatim}

\makeatletter

   \@addtoreset{equation}{section}
\makeatother

\begin{document}

\vskip 12mm

\begin{center} 
{\Large \bf  Vertex Operators for Irregular Conformal Blocks: Supersymmetric Case}
\vskip 10mm
{ \large  Dmitri Polyakov$^{a,b,}$\footnote{email:polyakov@scu.edu.cn;polyakov@sogang.ac.kr}
and  Chaiho Rim$^{c,}$\footnote{email:rimpine@sogang.ac.kr} }

\vskip 8mm
$^{a}$ {\it  Center for Theoretical Physics, College of Physical Science and Technology}\\
{\it  Sichuan University, Chengdu 6100064, China}\\
\vskip 2mm

$^{b}$ {\it Institute of Information Transmission Problems (IITP)}\\
{\it  Bolshoi Karetny per. 19/1, Moscow 127994, Russia}\\
\vskip 2mm
$^{c}$ {\it Department of Physics and Center for Quantum Spacetime}\\
{\it Sogang University, Seoul 121-742, Korea}

\end{center}

\vskip 15mm

\begin{abstract}

We  construct supersymmetric irregular vertex operators of arbitrary rank, appearing in the colliding
limit of primary fields.
We find that the structure of the supersymmetric irregular vertices differs significantly from the bosonic case:
upon supersymmetrization, the irregular operators are no longer the eigenstates of positive Virasoro and $W_N$ generators
but block-diagonalize them.
We relate the block-diagonal structure of the irregular vertices to contributions of the Ramond sector to the colliding limit.

\end{abstract}

\vskip 12mm

\setcounter{footnote}{0}

\section{\bf Introduction} 

Primary vertex operators in two-dimensional conformal field theory
are the objects playing  an important role in the AGT conjecture \cite{AGT},
connecting regular Liouville conformal blocks to Nekrasov's partition function
\cite{Nekrasov}
for  $N=2$
supersymmetric gauge theories in four dimensions.
These theories, however, admit a special case
with non-trivial IR fixed point, described by
Argyres-Douglas  theory \cite{AD, APSW}.
This class of theories does not allow marginal deformations
and is  described in terms of colliding limit of the primary 
vertex operators.  The operator of rank $q$, obtained from the colliding limit
\cite{EM_2009,GT_2012},
generates a irregular state of rank $q$ 
when applied to the vacuum. 
The irregular states of rank $q$ are annihilated by positive Virasoro generators $L_k$ with $k > 2q$
and are simultaneous eigenstates of  
$L_k$ with $q \le k \le 2q$.
These irregular states are called Gaiotto states \cite{G_2009}
or Whittaker states \cite{Whittaker}.  
In the generalizations of Liouville theory to Toda field theories,
the irregular states also possess $W_N$ symmetries, being simultaneous eigenstates of positive $W_N$ generators.
For example, in the simplest case of two fields the irregular states are  eigenvectors of positive $W_3$ generators
${(W_3)}_k$ with $2q\leq{k}\leq{3q}$.  For higher numbers of components, higher $W_N$ symmetries are switched on as well,
and the irregular states are the eigenstates of ${(W_N)}_k$ with $N(q-1)\leq{k}\leq{Nq}$

In our  previous paper \cite{RP} we have performed explicit construction of the irregular vertex operators
describing these states.
In general, the irregular vertex operators describing the states of rank $q$ for $n$-component Toda field
${\vec{\phi}}=(\phi_1,...\phi_n)$  are given by

\begin{eqnarray}
U_{n,q}=e^{\sum_{k=0}^q{\vec{\alpha}}_k\partial^k{\vec{\phi}}}
\end{eqnarray}
The Virasoro eigenvalues of $U_{n,q}$ are bilinear in the components of ${\vec{\alpha}}$, while the 
$W_N$ eigenvalues are polynomials of degree $N$ in $\alpha$ (see also \cite {CRZ_2015, nagoya, lefloch} for related reference).

The objects of the type (1.1)  generalize regular vertex operators in Toda theories and
in bosonic string theory and are in fact of interest far beyond their 
relevance to gauge theories in four dimensions.
First of all, such objects have to appear naturally in $AdS/CFT$ correspondence, being the AdS
 string duals  of composite operators on the CFT side.
For example, the operators of the type $T^q{\sim}T_{m_1n_1}...T_{m_qn_q}$
(with $T_{mn}$ being the energy-momentum tensor in super Yang-Mills theory) have to correspond
to the colliding limit of $q$ gravitons, or the irregular vertex of rank $q$ in AdS string theory.
At the same time, the objects of this type should be dual to higher spin fields with mixed symmetries
in $AdS$ higher spin gravity. Unlike regular vertex operators,
these objects are not in BRST cohomology and are off-shell. Nevertheless, they are completely meaningful
in string field theory (SFT) context.
On the other hand, the worldsheet correlators
of $U_{n,q}$ can be understood as generating functions for the higher spin interactions in string theory
since the derivatives $H^{m_1...m_s}{{\partial}\over{\partial_{\alpha_{n_1}}^{m_1}}}...
{{\partial}\over{\partial_{\alpha_{n_s}}^{m_s}}}U_{n,q}|_{{\vec{\alpha}}_1...{\vec{\alpha}}_n=0}$
describe on-shell higher spin vertex operators with momentum ${\vec{\alpha}}_0$ and masses 
$m\sim(\sum_{j=1}^s{n_j})^{{1\over2}}$, making the irregular vertex operators (1.1) natural candidates for
SFT solutions describing non-perturbative higher spin backgrounds.

Since the irregular states are typically of  relevance to models with supersymmetries,
it is naturally of interest to find the supersymmetric extensions of (1.1).
This turns out to be a major challenge in  conventional matrix model approach
and , so far, very little  progress has been made in describing supersymmetrizing the colliding limit.
The purpose of this work is to elaborate on this problem using the 
superspace extension of  the vertex operator formalism (1.1).
The results that we find are somewhat surprising. It turns out that the states, emerging in the supersymmetric
colliding limit, no longer diagonalize Virasoro generators. 
Instead, the positive Virasoro generators turn out to be block-diagonal
in their basis, with the multiplicities related to the ranks of the blocks.
Such a situation is familiar in logarithmic conformal field theories (LCFT) where the Hamiltonian
$L_0$ is also well-known to be a Jordan matrix. 
 The Hamiltonian being a Jordan  
block leads to appearance of the logarithmic partners and modified conformal symmetry  constraints
on correlation functions involving these partners. Solving these constraints leads to emergence
of logarithms in the correlators.
The supersymmetric extension of the vertices (1.1) thus, in a sense, may give the irregular analogues
of logarithmic operators in LCFT in the interacting case, although we will not elaborate on this question in the present letter.

The rest of the paper is organized as follows. In the  Section 2 we perform the supersymmetrization
of the operators (1.1) using the superspace approach.
Next, we study the properties of the supersymmetric irregular vertices under 
Virasoro transformations. In the Section 3 we discuss the reasons behind the appearance of the block-diagonal structure
for the irregular vertices, with the positive super Virasoro generators having the form of Jordan blocks in their basis.
We relate this structure to contributions of operators from the Ramond  sector to the colliding limit,
with the multiplicities of eigenvalues of the Jordan blocks connected to various possible
structures of these contributions.

\section{\bf  Superconformal Properties of  Irregular vertices}

The vertex operators (1.1) can be supersymmetrized in a straightforward and manifest way
by replacing the fields by the superfields and covariantizing the derivatives:
\begin{eqnarray}
{\vec{\phi}}(z)\rightarrow{\vec{\phi}}(z,\theta)={\vec{\phi}}(z)+\theta{\vec{\psi}}
\nonumber \\
\partial\equiv\partial_z\rightarrow{D}=\theta\partial_z+\partial_\theta
\end{eqnarray}
and expanding in $\theta$.
Using

\begin{eqnarray}
D^{2k}{\vec{\phi}}(z,\theta)=\partial^k{\vec{\phi}}+\theta\partial^k{\vec{\psi}}
\nonumber \\
D^{2k+1}{\vec{\phi}}(z,\theta)=\partial^k{\vec{\psi}}+\theta\partial^{k+1}{\vec{\phi}}
\nonumber \\
k=0,1,...
\end{eqnarray}
one easily finds
\begin{eqnarray}
W_{q+{1\over2}}=e^{\sum_{n=0}^{2q+1}{\vec{\alpha}}_nD_\theta^n{\vec{\phi}}(z,\theta)}
=e^{\sum_{n=0}^q{\vec{\alpha}}_{2n}\partial^n{\vec{\phi}}}
\nonumber \\
\times
\prod_{k=0}^q(1+{\vec{\alpha}}_{2k+1}\partial^k{\vec{\psi}})(1+\theta
(\sum_{n=0}^q{\vec{\alpha}}_{2n}\partial^n{\vec{\psi}}-{\vec{\alpha}}_{2n+1}\partial^{n+1}{\vec{\phi}})
\nonumber \\
W_q=e^{\sum_{n=0}^{2q}{\vec{\alpha}}_nD_\theta^n{\vec{\phi}}(z,\theta)}
=e^{\sum_{n=0}^q{\vec{\alpha}}_{2n}\partial^n{\vec{\phi}}}
\prod_{k=0}^{q-1}(1+{\vec{\alpha}}_{2k+1}\partial^k{\vec{\psi}})
\nonumber \\
\times
(1+\theta(\alpha_{2q}\partial^{q}\psi+
\sum_{n=0}^{q-1}{\vec{\alpha}}_{2n}\partial^n{\vec{\psi}}-{\vec{\alpha}}_{2n+1}\partial^{n+1}{\vec{\phi}}))
\end{eqnarray}
for the odd and even ranks respectively.
The ${\vec{\alpha}}_k$ coefficients are Grassmann even for even k's and Grassmann odd for the odd k's.
Clearly, the supersymmetrization reduces the total rank of the operators by half.
In particular, $W_{1\over2}$ is just a combination of regular vertex operators with different conformal dimensions:

\begin{eqnarray}
W_{1\over2}=e^{{\vec{\alpha}}{\vec{\phi}}}(1+{\vec{\beta}}{\vec{\psi}}
+\theta({\vec{\alpha}}{\vec{\psi}}-{\vec{\beta}}\partial{\vec{\phi}}))
\end{eqnarray}
or simply
\begin{eqnarray}
U_{1\over2}=\int{d\theta}W_1=e^{{\vec{\alpha}}{\vec{\phi}}}
({\vec{\alpha}}{\vec{\psi}}-{\vec{\beta}}\partial{\vec{\phi}})
\end{eqnarray}
upon $\theta$-integration.
Unlike the irregular bosonic vertices,
the operators (2.4)-(2.5) are not by themselves the eigenvectors of positive super Virasoro generators.
For example, $W_{1\over2}$ is annihilated  by $L_n$ for $n\geq{2}$, however the action of $L_1$ on $W_{1\over2}$
gives
\begin{eqnarray}
L_1W_{1\over2}=\theta{e^{{\vec{\alpha}}{\vec{\phi}}}}{\vec{\beta}}{\vec{\psi}}\equiv\theta{\tilde{W}}_{1\over2}
\nonumber \\
L_1^2W_{1\over2}=0
\end{eqnarray}
implying that $L_1$ acts on $(W_{1\over2},{\tilde{W}}_{1\over2})$
as a block-diagonal Jordan matrix  with $(W_{1\over2},{\tilde{W}}_{1\over2})$ having
eigenvalue zero and multiplicity 2 (for brevity and simplicity, here and elsewhere we set the background charge $q$
to zero; the generalization to $q\neq{0}$ is, however, straightforward) 
Next, consider the case of $W_1$.
In components, one has
\begin{eqnarray}
W_{1}=e^{{\vec{\alpha}}{\vec{\phi}}+{\vec{\gamma}}\partial{\vec{\phi}}}
(1+{\vec{\beta}}{\vec{\psi}})(1+\theta({\vec{\alpha}}{\vec{\psi}}
-{\vec{\beta}}\partial{\vec{\phi}}+{\vec{\gamma}}\partial{\vec{\psi}}))
\end{eqnarray}
This operator block-diagonalizes $L_1$ and $L_2$.
Acting with $L_2$, one obtains
\begin{eqnarray}
L_2W_{1}
=-{1\over2}\gamma^2{W_1}
+{1\over2}\theta{\vec{\beta}}{\vec{\gamma}}e^{{\vec{\alpha}}{\vec{\phi}}+{\vec{\gamma}}\partial{\vec{\phi}}}
\equiv-{1\over2}\gamma^2{W_1}+{\tilde{W}}_1^{(1)}
\nonumber \\
L_2{\tilde{W_{1}^{(1)}}}=-{1\over2}\gamma^2{\tilde{W_{1}^{(1)}}}
\end{eqnarray}
implying the block-diagonal structure of $L_2$ acting on $W_1$ with eigenvalue $-{1\over2}\gamma^2$ and multiplicity 2.
Acting with $L_1$, one obtains
\begin{eqnarray}
L_1W_{1}
=-{\vec{\alpha}}{\vec{\gamma}}{W_1}+({\vec{\gamma}}{\vec{\psi}}+\theta{\vec{\alpha}}{\vec{\beta}})
e^{{\vec{\alpha}}{\vec{\phi}}+{\vec{\gamma}}\partial{\vec{\phi}}}
\equiv
-{\vec{\alpha}}{\vec{\gamma}}{W_1}+{\tilde{W}}_1^{(2)}
\nonumber \\
L_1{\tilde{W}}_1^{(2)}=-{\vec{\alpha}}{\vec{\gamma}}{\tilde{W}}_1^{(2)}
\end{eqnarray}
implying the block-diagonal structure of $L_1$ acting on $W_1$ with eigenvalue $-{\vec{\alpha}}{\vec{\gamma}}$ and multiplicity 
2.
For the $W_{3\over2}$ case , the structure becomes more diverse.

The operator in components is
\begin{eqnarray}
W_{3\over2}=e^{{\vec{\alpha}}{\vec{\phi}}+{\vec{\gamma}}\partial{\vec{\phi}}}
(1+{\vec{\beta}}{\vec{\psi}})(1+{\vec{\lambda}}\partial{\vec{\psi}})
(1+\theta({\vec{\alpha}}{\vec{\psi}}+{\vec{\beta}}\partial{\vec{\phi}}
+{\vec{\gamma}}\partial{\vec{\psi}}
+{\vec{\lambda}}\partial^2{\vec{\phi}}))
\end{eqnarray}
It block-diagonalizes the Virasoro generators $L_1$, $L_2$ and $L_3$
with multiplicities $4,3$ and $2$ according to:

\begin{eqnarray}
L_1W_{3\over2}=-{\vec{\alpha}}{\vec{\gamma}}W_{3\over2}+{\tilde{W}}_{3\over2}^{1,1}
\nonumber \\
L_1
{\tilde{W}}_{3\over2}^{1,1}=-{\vec{\alpha}}{\vec{\gamma}}{\tilde{W}}_{3\over2}^{1,1}+{\tilde{W}}_{3\over2}^{1,2}
\nonumber \\
L_1
{\tilde{W}}_{3\over2}^{1,2}=-{\vec{\alpha}}{\vec{\gamma}}{\tilde{W}}_{3\over2}^{1,2}+{\tilde{W}}_{3\over2}^{1,3}
\nonumber \\
L_1
{\tilde{W}}_{3\over2}^{1,3}=-{\vec{\alpha}}{\vec{\gamma}}{\tilde{W}}_{3\over2}^{1,3}
\end{eqnarray}
for $L_1$,
\begin{eqnarray}
L_2W_{3\over2}=-{1\over2}{\gamma^2}W_{3\over2}+{\tilde{W}}_{3\over2}^{2,1}
\nonumber \\
L_2
{\tilde{W}}_{3\over2}^{2,1}=-{1\over2}{\gamma^2}{\tilde{W}}_{3\over2}^{2,1}+{\tilde{W}}_{3\over2}^{2,2}
\nonumber \\
L_2
{\tilde{W}}_{3\over2}^{2,2}=-{1\over2}{\gamma^2}{\tilde{W}}_{3\over2}^{2,2}
\end{eqnarray}
for $L_2$ and 
\begin{eqnarray}
L_3W_{3\over2}={\tilde{W}}_{3\over2}^{3,1}
\nonumber \\
L_3{\tilde{W}}_{3\over2}^{3,1}=0
\end{eqnarray}
for $L_3$ with
\begin{eqnarray}
{\tilde{W}}_{3\over2}^{1,1}=
e^{{\vec{\alpha}}{\vec{\phi}}+{\vec{\gamma}}\partial{\vec{\phi}}}
\times\lbrace
(1+{\vec{\beta}}{\vec{\psi}})(1+{\vec{\lambda}}\partial{\vec{\psi}})
({\vec{\alpha}}{\vec{\beta}}-2{\vec{\lambda}}\partial{\vec{\phi}}
-{\vec{\gamma}}{\vec{\psi}})\theta
\nonumber \\
+(1+({\vec{\beta}}+{\vec{\lambda}}){\vec{\psi}})
(1+\theta({\vec{\alpha}}{\vec{\psi}}
+{\vec{\beta}}\partial{\vec{\phi}}
+{\vec{\gamma}}\partial{\vec{\psi}}
+
{\vec{\lambda}}\partial^2{\vec{\phi}}))\rbrace
\nonumber \\
{\tilde{W}}_{3\over2}^{1,2}=
2e^{{\vec{\alpha}}{\vec{\phi}}+{\vec{\gamma}}\partial{\vec{\phi}}}
\times\lbrace
(1+{\vec{\beta}}{\vec{\psi}})(1+{\vec{\lambda}}\partial{\vec{\psi}})
{\vec{\alpha}}{\vec{\lambda}}
\nonumber \\
+
(1+({\vec{\beta}}+{\vec{\lambda}}){\vec{\psi}})(
{\vec{\alpha}}{\vec{\beta}}-{\vec{\lambda}}\partial{\vec{\phi}}-{\vec{\gamma}}{\vec{\psi}})
\rbrace\theta
\nonumber \\
{\tilde{W}}_{3\over2}^{1,3}=
4e^{{\vec{\alpha}}{\vec{\phi}}+{\vec{\gamma}}\partial{\vec{\phi}}}
(1+({\vec{\beta}}+{\vec{\lambda}}){\vec{\psi}}){\vec{\alpha}}{\vec{\lambda}}\theta
\nonumber \\
{\tilde{W}}_{3\over2}^{2,1}=
e^{{\vec{\alpha}}{\vec{\phi}}+{\vec{\gamma}}\partial{\vec{\phi}}}
\times\lbrace
(1+{\vec{\beta}}{\vec{\psi}})(1+{\vec{\lambda}}\partial{\vec{\psi}})
({\vec{\gamma}}{\vec{\beta}}+2{\vec{\alpha}}{\vec{\lambda}})\theta
\nonumber \\
-{1\over2}{\vec{\beta}}{\vec{\lambda}}
(1+\theta({\vec{\alpha}}{\vec{\psi}}+{\vec{\beta}}\partial{\vec{\phi}}
{\vec{\gamma}}\partial{\vec{\psi}}
{\vec{\lambda}}\partial^2{\vec{\phi}}))+
{1\over2}((1+{\vec{\lambda}}\partial{\vec{\psi}}){\vec{\beta}}{\vec{\gamma}}
-(1+{\vec{\beta}}{\vec{\psi}}){\vec{\alpha}}{\vec{\lambda}})\theta\rbrace
\nonumber \\
{\tilde{W}}_{3\over2}^{2,2}=
-e^{{\vec{\alpha}}{\vec{\phi}}+{\vec{\gamma}}\partial{\vec{\phi}}}
({\vec{\beta}}{\vec{\gamma}}+2{\vec{\alpha}}{\vec{\lambda}}){\vec{\beta}}{\vec{\lambda}}
\theta
\nonumber \\
{\tilde{W}}_{3\over2}^{3,1}=
-2e^{{\vec{\alpha}}{\vec{\phi}}+{\vec{\gamma}}\partial{\vec{\phi}}}
{\vec{\gamma}}{\vec{\lambda}}\theta
\end{eqnarray}

Accordingly, the Jordan block constraints for $W_{3\over2}$ are

\begin{eqnarray}
(L_1+{\vec{\alpha}}{\vec{\gamma}})^4W_{3\over2}=0
\nonumber \\
(L_2+{1\over2}\gamma^2)^3W_{3\over2}=0
\nonumber \\
L_3^2W_{3\over2}=0
\end{eqnarray}

It is straightforward to check that the same pattern holds for the higher $q$'s as well.
At half integer orders, the operators $W_{q+{1\over2}}$
block-diagonalize $L_n$ with $q\leq{n}\leq{2q+1}$
with the eigenvalues given by
\begin{eqnarray}
\lambda_n=-{1\over2}\sum_{0{\leq}k_1;k_2\leq{q};k_1+k_2=n}k_1!k_2!{\vec{\alpha}}_{k_1}{\vec{\alpha}}_{k_2}
\end{eqnarray}
for $n\leq{2q}$ and $\lambda_{2q+1}=0$.
The Jordan block conditions are
\begin{eqnarray}
(L_n-\lambda_n)^{m(n,q)}W_{q+{1\over2}}=0
\end{eqnarray}
with the multiplicities ranging from

\begin{eqnarray}
m(2q+1,q)=1+{\lbrack}{2\over{q}}+{{q+1}\over2}\rbrack
\end{eqnarray}
for the highest Virasoro operator $L_{2q+1}$
to $m(q,q)\geq{\lbrack}{{q}\over4}+{5\over8}+{{13}\over{8q}}\rbrack$ for the lowest Virasoro generator $L_q$.
Likewise, at the integer orders the $W_q$ operators  block-diagonalize the Virasoro generators
from $L_q$ to $L_{2q}$ with the similar eigenvalues $\lambda_n$ (2.16) and with the multiplicities ranging from
\begin{eqnarray}
m(2q,q)={\lbrack}{{q+1}\over{2q}}\rbrack+q
\end{eqnarray}
for $L_q$ and $m(q,q){\geq}{q\over2}+{\lbrack}{{q+1}\over{4q}}\rbrack$ for $L_{2q}$.
The analysis of $W_q$ behaviour under the supersymmetry transformations
by the supersurrent modes is performed analogously.
The irregular vertices  block-diagonalize the supercurrent modes from $G_{q-{1\over2}}$ up to
$G_{2q-{1\over2}}$  for integer ranks and $G_{2q+{1\over2}}$ for half-integer ranks. 
This  concludes the analysis of the CFT properties of supersymmetric irregular vertex operators.

\section{\bf Colliding Limit and Block-Diagonal Structure}

In the bosonic case, the irregular vertex operators with the structure
$W_q=e^{\sum_n=0^q{\vec{\alpha}}_n\partial^n{\vec{\phi}}}$ 
appear in the colliding limit of $q$ regular vertex operators as a result of 
simultaneous normal ordering which structure
is rather complicated.
Namely,
for $N$ colliding operators sitting at points $z_1,...,z_N$
one has
for N vertices at $z_1,...z_N$ around $z_1$
\begin{align}
W_N=\lim_{z_2,...z_N\rightarrow{z_1}}e^{\alpha_1\phi}(z_1)...e^{\alpha_N}(z_N)
\nonumber \\
=\prod_{p=2}^N(z_{k1})^{-\alpha_p(\alpha_1+...\alpha_{p-1})}
\sum_{n_1,...,n_{N-1}}\sum_{k_1,...k_{N-2}}\sum_{q_1....q_{N-2}}
(z_{21})^{n_1}(z_{31})^{n_2-k_1}...
\nonumber \\
\times
(z_{N1})^{n_{N-1}-k_{N-2}}                                                                                                                     
\prod_{j=1}^{N-1}
\lambda_{\lbrace{n,k,q}\rbrace}
:B^{(n_j-q_j(k_1,...k_{N-1}))}_{\alpha_{j+1}}
e^{(\alpha_1+...+\alpha_N)\phi}:(z_1)
\label{eq:N-fusion}
\end{align}
with
the $q$-numbers satisfying
\begin{eqnarray}
\sum_{j=1}^{N-2}k_j=\sum_{j=1}^{N-2}q_j
\end{eqnarray}
and $\lambda_{\lbrace{n,k,q}\rbrace}$ are some constants.
Here $z_{ij}= z_i-z_j$ and
 $B^{(n)}_\alpha$ are the normalized Bell polynomials in the derivatives
of $\phi$  defined as \cite{polyakov_2015}
\begin{eqnarray}
B^{(n)}_\alpha=\sum_{p=1}^n\alpha^p\sum_{n|k_1...k_p}
{{\partial^{k_1}\phi...\partial^{k_p}\phi}\over{k_1!q_{k_1}!...k_p!q_{k_p}!}}\,.
\end{eqnarray}
Here the sum is taken over the ordered length $p$ partitions
of $n$ $(1\leq{p}\leq{n})$: $n=k_1+...+k_p$;
$k_1{\leq}k_2...\leq{k_p}$
and $q_{k_j}$ is the multiplicity of an element $k_j$ in the partition.
The irregular operators, 
appearing as a result of the summation, diagonalize positive Virasoro generators,
with the derivatives of fields in the exponents appearing as a result of summing over
all orders of $z_{ij}$, involving the exponents multiplied by Bell polynomials of various
degrees.  The Bell polynomials (times exponents) form a basis which is somewhat inconvenient
to describe the colliding limit, therefore, the relation between complicated summation
(3.1) and a relatively simple structure of the resulting irregular vertex operator (1.1) is not straightforward
to obtain. Nevertheless, the colliding limit may be useful to illustrate the difference between the bosonic
case (where the Virasoro operators are diagonal) and the  supersymmetric case, where they become the Jordan blocks
(with the eigenvalues, inherited from the irregular bosonic  vertices, except for the half-integer case,
where an extra block-diagonal generator $L_{2q+1}$ with the zero eigenvalue appears as well).
The important difference between these two cases is related to the fact that, the supersymmetric irregular vertex 
operators involve the colliding limit of the regular vertices from the Ramond sector
involving $\sigma_{\pm}$, satisfying
\begin{eqnarray}
\sigma_{+}(z)\sigma_{-}(w)\sim(z-w)^{3\over{8}}\psi
\nonumber \\
\sigma_{+}(z)\sigma_{+}(w)\sim(z-w)^{-{1\over{8}}}
\nonumber \\
\sigma_{-}(z)\sigma_{-}(w)\sim(z-w)^{-{1\over{8}}}
\end{eqnarray}
The simplest supersymmetric vertex,
$W_{1\over2}$, appears as a result of a collision of
spin operator multiplied by exponent of super Liouville field
with the spin operator without exponents:
\begin{eqnarray}
\lim_{z\rightarrow{w}}:\sigma_{\pm}e^{\alpha\phi}(z): :\sigma_{\pm}:(w)
\end{eqnarray}
(or equivalently, OPE expansion and normal ordering  around the midpoint may be considered)
In such a collision, the exponent does not develop irregularity
(appearance of derivatives of $\phi$) because all the Bell polynomials
$B_{\alpha}$ appearing in the collision will multiply by the exponent with the same $\alpha$,
producing total derivatives of $e^{\alpha\phi}$ that will sum up to the exponent at the location of the
normal ordering.
The higher ranks, such as $W_1$ or $W_2$, admit more possibilities in the colliding limit.
For example,  one can either collide $e^{\alpha\phi}\sigma_{\pm}$ at $z_1$ with
$e^{\beta\phi}\sigma_{\pm}$ at $z_2$, or collide the same two exponents
with extra insertions of the spin operators $\sigma_{\pm}$, say, at $z_3$ and $z_4$ with 
the collision structure $\sigma_\pm{e^{\alpha\phi}}-\sigma_\pm{e^{\beta\phi}}
-\sigma_\pm-\sigma_\pm$.
In the absense of supersymmetry, the insertion number of free $\sigma$'s can be arbitrary;
in the supersymmetric case the number is limited. It is precisely this ``ambiguity'' that 
stands behind the appearance of multiplicities of eigenvalues and, as a result, the super Virasoro generators
become the Jordan blocks.
\section{\bf Conclusion}

In this letter we have constructed the supersymmetric extension of irregular vertex operators,
which correlation functions describe irregular supersymmetric conformal blocks.
We found that the superconformal structure of the irregular vertices differs significantly from the bosonic case,
with the supersymmetric irregular operators block-diagonalizing the positive Virasoro generators.
The block-diagonal form of the positive Virasoro generators  in the supersymmetric case is related to contributions
of the operators from the Ramond sector to the colliding limit.
Although in this letter we only considered the Virasoro symmetry in details, it is straightforward to check
that, in case of the multiple (Toda) fields, the same block-diagonal structure emerges when the 
irregular vertices are transformed by the positive modes of the $W_N$-currents.
It will be important to understand the impact of the structures, observed in this work, in the language of the matrix model approach such as in \cite{EM_2009,CRZ_2015,NR_2012,CRZ_2014,CR_2015} 
and their relevance to Seiberg-Witten curves in related supersymmetric gauge theories in four dimensions.
As we also mentioned, the irregular vertex operators appear to be natural objects  in background-independent string field theory,
being natural blocks for solutions describing both nonperturbative higher-spin backgrounds, as well as cosmological
backgrounds, such as the noncommutative rolling tachyons. We hope to be able to address these questions soon in the 
future papers.

\begin{center}
{\bf Acknowledgements}
\end{center}

The authors acknowledge the support of this work by the National Research Foundation of Korea(NRF)
 grant funded by the Korea government(MSIP) (NRF-2014R1A2A2A01004951) and by the National Natural 
Science Foundation of China under grant 11575119.


\end{document}